\documentclass[aps,prb,twocolumn,superscriptaddress]{revtex4-1}

\usepackage{amssymb}
\usepackage{color}
\usepackage{bm}
\usepackage{epsfig}
\usepackage[latin1]{inputenc}
\usepackage{graphicx}
\usepackage{dcolumn}

\usepackage{setspace}

\begin{document}


\title{Pressure induced amorphization and collapse of magnetic order \\
in type-I clathrate Eu$_8$Ga$_{16}$Ge$_{30}$}

\author{J. R. L. Mardegan}
\affiliation{Instituto de F\'isica ``Gleb Wataghin", Universidade Estadual de Campinas, Campinas, S\~ao Paulo 13083-859  Brazil}
\affiliation{Advanced Photon Source, Argonne National Laboratory, Argonne, Illinois 60439 USA}

\author{G. Fabbris}
\affiliation{Advanced Photon Source, Argonne National Laboratory, Argonne, Illinois 60439 USA}
\affiliation{Department of Physics, Washington University, St. Louis, Missouri 63130 USA}

\author{L. S. I. Veiga}
\affiliation{Instituto de F\'isica ``Gleb Wataghin", Universidade Estadual de Campinas, Campinas, S\~ao Paulo 13083-859 Brazil}
\affiliation{Laborat\'orio Nacional de Luz S\'incrotron, Campinas, S\~ao Paulo 13083-970 Brazil}

\author{C. Adriano}
\altaffiliation[Present address: ]{University of Illinois, Chicago, Illinois 60607 USA}
\affiliation{Instituto de F\'isica ``Gleb Wataghin", Universidade Estadual de Campinas, Campinas, S\~ao Paulo 13083-859 Brazil}

\author{M. A. Avila}
\affiliation{CCNH, Universidade Federal do ABC (UFABC), Santo Andr\'e, S\~ao Paulo 09210-580 Brazil}

\author{D. Haskel}
\affiliation{Advanced Photon Source, Argonne National Laboratory, Argonne, Illinois 60439 USA}

\author{C. Giles}
\email[]{giles@ifi.unicamp.br}
\altaffiliation[Temporary address: ]{Advanced Photon Source, Argonne National Laboratory, Argonne, Illinois 60439 USA}
\affiliation{Instituto de F\'isica ``Gleb Wataghin", Universidade Estadual de Campinas, Campinas, S\~ao Paulo 13083-859 Brazil}

\date{\today}


\begin{abstract}
We investigate the low temperature structural and electronic properties of the type-I clathrate Eu$_8$Ga$_{16}$Ge$_{30}$ under pressure using x-ray powder diffraction (XRD), x-ray absorption near-edge structure (XANES) and x-ray magnetic circular dichroism (XMCD) techniques. The XRD measurements reveal a transition to an amorphous phase above 18~GPa. Unlike previous reports on other clathrate compounds, no volume-collapse is observed prior to the crystalline-amorphous phase transition which takes place when the unit cell volume is reduced to 81$\%$ of its ambient pressure value. Fits of the pressure-dependent relative volume to a Murnaghan equation of state (EOS) yield a bulk modulus $B_0=65\pm3$~GPa and a pressure derivative $B'_0=3.3\pm0.5$. The Eu L$_2$-edge XMCD data shows quenching of the magnetic order at the crystalline-amorphous phase transition. The XANES spectra indicate the persistence of Eu$^{2+}$ valency state up to 22~GPa, therefore the suppression of XMCD intensity is due to the loss of magnetic order as a result of frustrated exchange interactions in the amorphous phase, and not due to quenching of local moments. When compared with other clathrates, the results point to the importance of guest ion-cage interactions in determining the mechanical stability of the framework structure and the critical pressure for amorphization. Finally, the crystalline structure is not found to recover after pressure release, resulting in a novel amorphous material that is at least metastable at ambient pressure and temperature.

\end{abstract}



\maketitle


\section{INTRODUCTION}

Materials with enhanced thermoelectric properties are needed fueled partly by new developments in energy conversion from nanoscale and nanostructured thermoelectric materials.\cite{Szczech_JMC_2011} The thermoelectric figure of merit of materials is quantified by their dimensionless quantity $ZT = S^2 T / (\kappa \rho)$, where $S$, $T$, $\rho$ and $\kappa$ are the Seebeck coefficient, temperature, electrical resistivity and the thermal conductivity, respectively.\cite{Snyder_Nature_2008, Pei_Nature_2011} Significant effort has been directed at investigating complex intermetallic materials based on groups IV and V elements, due to their ability to form guest/host cage structures with enhanced phonon scattering, and consequent low thermal conductivity. Two representative families are the filled skutterudite compounds with general formula RM$_4$X$_{12}$ and the type-I clathrate compounds with general formula A$_8$X$_{46}$, in which host cages made of various combinations of X atoms from the aforementioned element groups are filled by R or A guest ions such as rare earth and alkaline earth elements. In addition to the extensively investigated enhanced thermoelectric properties~\cite{Sales_PRB_2001, Paschen_PRB_2001} these materials also display a myriad of complex phenomena, such as superconductivity,\cite{Meisner_Elsevier_1984} metal-insulator transitions,\cite{Sekine_PRL_1997} magnetic ordering,\cite{Paschen_PRB_2001, Meisner_JAP_1985, Torikachvili_JMMM_1986} Kondo insulator effects~\cite{Aeppli_CCMP_1992} and heavy fermion behavior.\cite{Maple_PB_1999, Jasper_SSC_1999, Gajewski_JP_1998, Takeda_PB_1999}

Among the type-I clathrates, Eu$_8$Ga$_{16}$Ge$_{30}$ is one of the most investigated and unique in that it presents full filling of the host cages with a rare earth element. This compound has potential in thermoelectric applications due to its behavior approaching the ``phonon glass, electron crystal'' (PGEC) concept.\cite{Slack_HT_1995} In addition, Eu$_8$Ga$_{16}$Ge$_{30}$ shows a variety of interesting properties such as anharmonic vibration of the Eu ions,\cite{Cohn_PRL_1999, Nolas_PRB_2000a, Nolas_PRB_2000b, Chakoumakos_JAC_2001, Takasu_PRB_2006} ferromagnetic ordering~\cite{Sales_PRB_2001, Paschen_PRB_2001} with Curie temperature $T_C \sim 35$ K, multiple (and as yet unresolved) magnetic structures below $T^* \sim$ 23 K,\cite{Hermann_PRL_2006, Onimaru_JPCS_2010, Phan_PRB_2011} enhanced magnetocaloric effect,\cite{Phan_PRB_2011, Phan_APL_2008, Srinath_APL_2006} development of magnetic polarons\cite{Rosa_PRB_2013} and structural dimorphism.\cite{Paschen_PRB_2001, Pacheco_PRB_2005, Bentien_PRB_2005}

\begin{figure}[!h]
\centering
\includegraphics[scale=1.05]{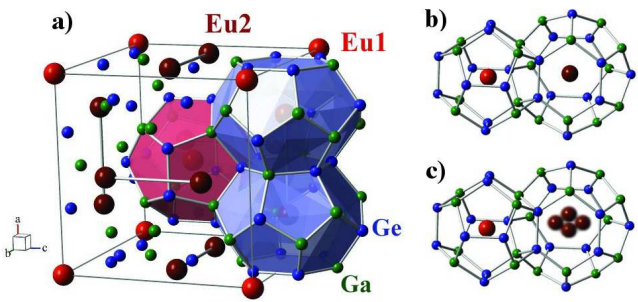}
\caption{(color online) Eu$_8$Ga$_{16}$Ge$_{30}$ structural representation. (a) Unit cell of the clathrate type-I Eu$_8$Ga$_{16}$Ge$_{30}$. The Eu atoms are shown at their two sites (Eu1 and Eu2) and inside the dodecahedral (X$_{20}$ - red polyhedra) and tetrakaidecahedral (X$_{24}$ - blue polyhedra) cages, respectively. The X$_{20}$ and X$_{24}$ polyhedra are shown in detail in (b) and (c), respectively. In (b) we show Eu1 at the 2$a$ and Eu2 at the 6$d$ Wyckoff positions. In (c) we show Eu1 at the 2$a$ and Eu2 at the 24$k$ Wyckoff positions. The Ga/Ge atoms (green and blue colors) are distributed at the 6$c$, 16$i$ and 24$k$ crystallographic positions.}
\label{fig:fig1}
\end{figure}

In its type-I clathrate phase, Eu$_8$Ga$_{16}$Ge$_{30}$ displays a cubic structure within the $P m \overline{3} n$ space group (No. 223). The unit cell presents two types of cages: six in the shape of tetrakaidecahedrons (X$_{24}$) and two smaller dodecahedral (X$_{20}$) ones (Figure~\ref{fig:fig1}(a)). The cages are formed by Ga and Ge ions distributed among the nonequivalent crystallographic sites 6$c$, 16$i$ and 24$k$.\cite{Sales_PRB_2001, Chakoumakos_JAC_2001} There is evidence that the Ga and Ge ions distribution is not entirely random, since the Ga ions preferentially occupy the 6$c$ and avoid the 16$i$ Wyckoff position.\cite{Bentien_ACIEE_2000, Zhang_APL_2002}

Inside the smaller X$_{20}$ polyhedron the Eu atom is hosted at the center of the cage (2$a$ Wyckoff position). On the other hand, the Eu ions inside the larger cages rattle among four equivalent off-center positions (24$k$ crystallographic sites) (Figure~\ref{fig:fig1}(b) and (c)).\cite{Sales_PRB_2001, Paschen_PRB_2001, Hermann_PRL_2006, Chakoumakos_JAC_2001} This rattling motion is of significant interest for the material's thermoelectric properties and it has been one of the main drivers behind investigations of this and other clathrate compounds.

Many of the clathrates physical properties are regulated by the guest-cage interaction and their relative sizes. Thus, applying pressure is a clean method to modify the guest-host interactions, and potentially tune these properties. Such effect is evident by the wide variety of new crystallographic, electronic and magnetic behaviors observed in these materials at high pressure. For instance, a polycrystalline Sr$_8$Ga$_{16}$Ge$_{30}$ sample with rather poor ambient pressure $ZT$ had its value strongly enhanced at 7~GPa.\cite{Meng_JAP_2001} The improved $ZT$ under pressure provides further valuable insight towards application-oriented design of thermoelectric materials. The isostructural compound Ba$_8$Ga$_{16}$Ge$_{30}$ was also investigated under pressure by x-ray diffraction and Raman spectroscopy up to 40~GPa,\cite{Kume_JAP_2012} through which a volume-collapse and anomalies in the spectral features were observed. This is consistent with several type-I clathrates studied under pressure exhibiting volume-collapse followed by a crystalline to amorphous transition at high pressure.~\cite{Kume_JAP_2012, Machon_PRB_2009, Yang_PRB_2006, Miguel_PRB_2002, Miguel_EL_2005, Tse_PRL_2002} Previous works on the Eu$_8$Ga$_{16}$Ge$_{30}$ clathrate under high pressure used Raman scattering to observe rattling vibrations of the Eu ion up to 6.7~GPa and 202~K,~\cite{Funahashi_JPCS_2012} and electrical resistivity and Hall coefficient measurements up to 11.4 GPa.~\cite{Umeo_JPCS_2012} An slight increase in $T_C$ and $T^*$ as a function of pressure and a decrease in resistivity as a result of an increase in carrier concentration were the main findings in these works. Consequently, the effect of high pressure in the electronic and crystallographic structure  of this compound remained elusive.

In this paper we report high pressure x-ray powder diffraction (XRD), x-ray absorption near-edge structure (XANES) and x-ray magnetic circular dichroism (XMCD) results on the type-I clathrate Eu$_8$Ga$_{16}$Ge$_{30}$ at low temperature, aiming to probe the pressure dependence of the various phenomena displayed by the Eu ions and the host cages. The data reveal an irreversible amorphization of the structure around 18 GPa. In contrast to other clathrates, no volume-collapse is found preceding the amorphization. Eu L$_2$ XMCD measurements show a sharp suppression of ferromagnetic moment in the 16-20 GPa range commensurate with the structural change. The absence of an Eu$^{3+}$ state signature in the XANES spectra points to the persistence of the 4$f^7$ local moment. Furthermore, the linear XMCD hysteresis loops indicate the loss of ferromagnetic order which, when associated to the XRD results, points to a pressure-induced paramagnetic amorphous state. The magnetic signal is not recovered on pressure release, consistent with an irreversible nature of the structural transition, which implies a novel amorphous phase in this system which is at least metastable at ambient pressure and temperature.


\section{EXPERIMENTAL DETAILS}

Single crystals of type-I Eu$_8$Ga$_{16}$Ge$_{30}$ were grown at IFGW/UNICAMP by a Ga self flux method similar to that previously detailed.\cite{Avila_PB_2006, Ribeiro_PM_2012} Selected crystals were ground and sieved through a 635 mesh, resulting in fine powder with grain sizes $\sim10~\mu$m. The XANES and XMCD spectra measurements at ambient pressure were performed in transmission mode with the sample mounted uniformly on tapes. High-pressure powder XRD and the XANES/XMCD measurements at ambient and high pressure were performed at beam lines 16-BM-D and 4-ID-D, respectively, of the Advanced Photon Source, Argonne National Laboratory. The XRD measurement was performed at 10 K and 30 K. The powder patterns were collected with an image plate detector (MAR345) with pixel size of 100~$\mu$m placed at 491~mm from the symmetric diamond anvil cell (DAC) (Princeton shops). The 2D images were integrated to provide intensity as a function of 2$\theta$ using the software FIT2D.\cite{Hammersley_HPR_1996} Due to the DAC limited angular scattering range ($\sim$18 degrees of scattering angle 2$\theta$), the beam was tuned to 29.2 keV in order to detect a significant number of Bragg peaks within this angular range. For the XRD measurements two full diamond anvils with 300~$\mu$m culet diameter were used. Re gaskets were pre-indented to 60~$\mu$m, and a 140~$\mu$m sample chamber was laser-drilled. He gas was used as pressure medium.\cite{Rivers_HPR_2008} Pressure was calibrated \textit{in situ} using ruby spheres with $\sim5~\mu$m diameter and a small amount of Au powder as standards.\cite{Syassen_HPR_2008} The Au peaks are marked with a * in the diffractograms.

High-pressure, low temperature ($T = 10$~K) XANES and XMCD measurements were performed in a transmission geometry at the Eu L$_2$ absorption edge (7617 eV).\cite{Haskel_HP_setup1, Haskel_HP_setup2} Some pressure points were measured at the Eu L$_3$ edge but due to the large number of Bragg reflection glitches in each X-ray absorption spectrum (XAS) we were only able to measure at the Eu L$_2$ edge. A pair of Si/Pd mirrors was used to focus and collimate the beam. Harmonic rejection was done both by the reflectivity cut off of the mirrors, and by detuning the double crystal monochromator. XMCD experiments were performed in helicity switching mode at 13.1 Hz, and the signal detected with a lock-in amplifier to increase the signal-to-noise ratio.\cite{Suzuki_JJAP_1998} An external magnetic field of H = 0.5 T was used to align the ferromagnetic domains. Measurements were repeated for opposite field directions to remove non-magnetic artifacts. All XANES and XMCD data were normalized to a jump of 1.0. A membrane-driven Copper-Beryllium (CuBe) diamond anvil cell (DAC) was used. Due to large absorption by the diamonds, a partially perforated anvil was opposite to a fully perforated anvil with a mini anvil on top.\cite{Haskel_HP_setup1, Haskel_HP_setup2} Independent experiments were carried out with culet diameters of 300, 450 and 600~$\mu$m. Pressure was calibrated \textit{in situ} with a ruby luminescence system,\cite{Syassen_HPR_2008} and silicone oil was used as pressure medium. Stainless steel gaskets pre-indented to 50-90~$\mu$m were used and the sample chamber drilled with an electrical discharge machine (EDM).


\section{EXPERIMENTAL RESULTS}

\subsection{XRD measurements}

\begin{figure}[!ht]
\centering
\includegraphics[trim=1.03cm 0.2cm 0.7cm 0.1cm, clip=true, totalheight=0.58 \textheight, angle=0]{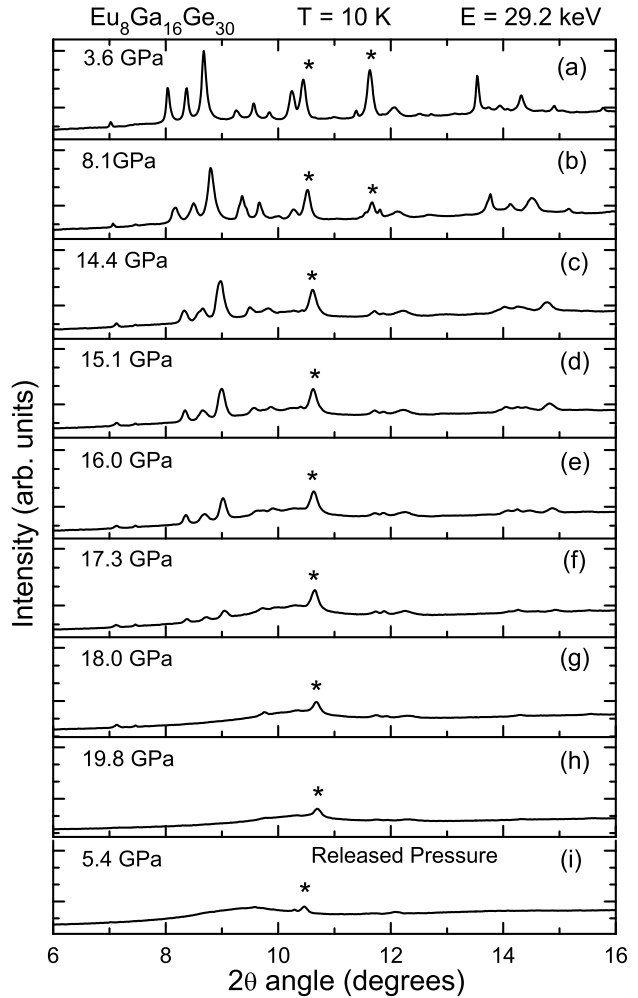}
\caption{XRD patterns of type-I clathrate Eu$_8$Ga$_{16}$Ge$_{30}$ at 10~K. Panels (a)-(h) show the diffraction patterns for increasing applied pressures. Panel (i) shows the diffraction pattern when the pressure is released after the sample reaches the crystal-to-amorphous transformation. The first three intense Bragg peaks (around 8 - 9 degrees) correspond to the (222), (320) and (321) reflections, respectively. The peaks marked with $*$ are due to Au powder.}
\label{fig:fig2}
\end{figure}

Figure~\ref{fig:fig2} shows selected XRD spectra obtained as a function of pressure at 10~K. No additional peaks or splittings are observed in the diffraction patterns to the highest pressure, indicating that the structure remains in the $P m \overline{3} n$ space group leading to the amorphization. At 19.8~GPa the sample Bragg peaks completely disappear and only a broad amorphous halo pattern is observed. This clearly shows that the type-I clathrate Eu$_8$Ga$_{16}$Ge$_{30}$ has a pressure induced crystalline-amorphous transition. Panel (i) in Figure~\ref{fig:fig2} shows the XRD pattern at~5.4~GPa after the decompression. XRD patterns were also measured after pressure release upon warming. The amorphous phase is also observed up to room temperature, indicating that this thermal energy is not sufficient to induce a recrystallization. A shift of the high-density amorphous pattern to lower 2$\theta$ values is observed for higher pressures. In addition, we have measured the XRD pressure evolution at 30~K (between $T^*$ and $T_C$) and we observed the same irreversible crystalline-amorphous transition at $\sim18$~GPa.

\begin{figure}[!hbt]
\centering
\includegraphics[trim=0.3cm 0.25cm 0.2cm 0.2cm, clip=true, totalheight=0.3\textheight, angle=0]{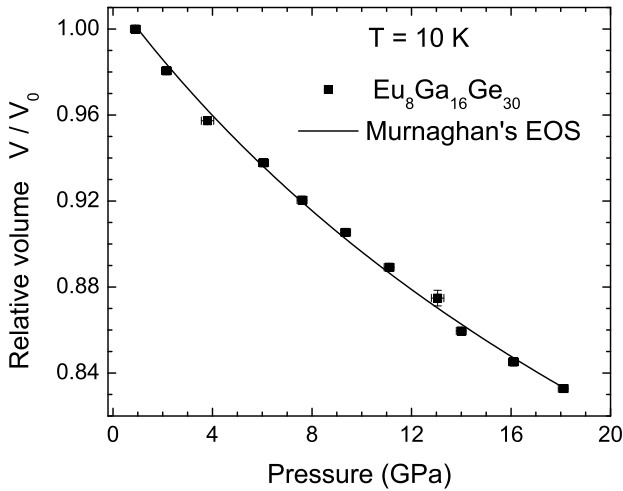}
\caption{Pressure-volume dependence of Eu$_8$Ga$_{16}$Ge$_{30}$ up to 18 GPa normalized by the volume at lower pressure (V$_0\sim$~1221.3~\AA$^3$ and P$_0\sim$~1~GPa). The experimental data were obtained from refinements of the XRD patterns measured at 10~K. The solid line represents the result of fitting by a Murnaghan equation of state. }
\label{fig:fig3}
\end{figure}

The evolution of the unit cell volume measured at 10~K normalized to the lowest pressure measured (V$_0\sim$~1221.3~\AA$^3$ and P$_0\sim$~1~GPa) is shown in Figure~\ref{fig:fig3}. The XRD measurements performed at 30~K (not shown here) showed the same pressure dependence and only a displacement smaller than 0.01~\AA~ in the lattice parameter for each pressure point. In contrast to other clathrates~\cite{Miguel_PRB_2002, Kume_JAP_2012, Machon_PRB_2009} no volume-collapse leading to the amorphization is observed. The unit cell volume was compressed to about 83$\%$ of the volume at 1~GPa at the onset of amorphization (81$\%$ compared to volume at ambient pressure). The solid line in Figure~\ref{fig:fig3} is a fit to a third-order Murnaghan equation of state (EOS).\cite{Murnaghan_PNAS_1944} This fit yields a bulk modulus $B_0=65\pm3$~GPa and a pressure derivative $B'_0=3.3\pm0.5$ for the Eu$_8$Ga$_{16}$Ge$_{30}$ structure.


\subsection{Absorption measurements}

\begin{figure}[!hbt]
\centering
\includegraphics[trim=0.75cm 0.2cm 0.1cm 0.1cm, clip=true, totalheight=0.3\textheight, angle=0]{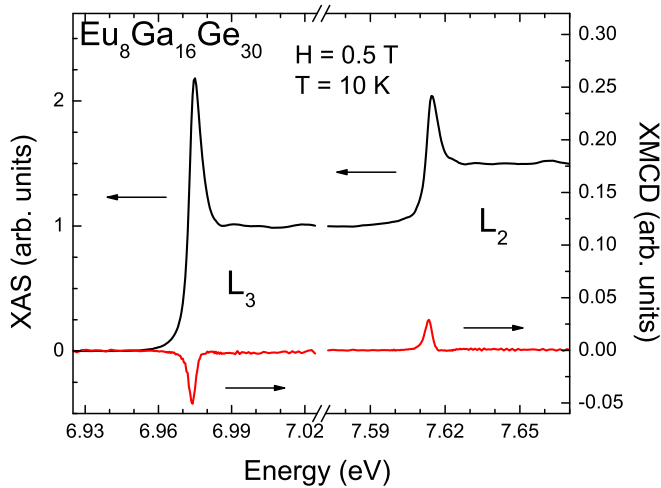}
\caption{(color online) Eu L$_{2,3}$ (a) XANES and (b) XMCD signal for the type-I clathrate Eu$_8$Ga$_{16}$Ge$_{30}$ measured at 10~K under an applied magnetic field of 0.5~T.}
\label{fig:fig4}
\end{figure}

The previously reported ambient pressure data for the clathrate Eu$_8$Ga$_{16}$Ge$_{30}$ obtained at Eu L$_{2,3}$ absorption edges~\cite{Krishnamurthy_PRB_2009} are in agreement with our XANES/XMCD data, whose normalized spectra measured at $T=10$~K and $H=0.5$~T are shown in Figure~\ref{fig:fig4}.

\begin{figure}[!hbt]
\centering
\includegraphics[trim=0.1cm 0.0cm 0.0cm 0.0cm, clip=true, totalheight=0.19\textheight, angle=0]{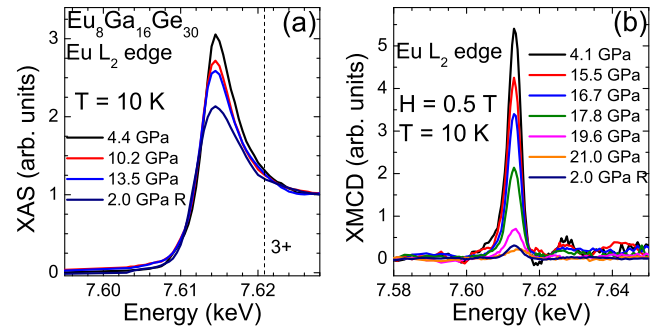}
\caption{(color online) Eu L$_2$-edge (a) XANES and (b) XMCD spectra for the type-I clathrate Eu$_8$Ga$_{16}$Ge$_{30}$ measured at 10 K and with an external applied magnetic field of 0.5 T at different pressures. The dashed vertical line marks the expected position of the Eu$^{3+}$ features. The XANES and XCMD spectra at 2.0 GPa was measured after releasing the pressure.}
\label{fig:fig5}
\end{figure}

The XANES and XMCD pressure evolution for the Eu L$_2$ edge are shown in Figures~\ref{fig:fig5}(a) and (b), respectively. The L$_2$ XANES spectra are sensitive to the europium ion valency, such that an additional spectral weight at 8~eV manifests as the signature of a 3+ component,\cite{Matsubayashi_PRB_2011, Souza-Neto_PRL_2012, Bi_PRB_2012} yet no such component is observed in our measured pressure range. In particular, no 4$f^7 \rightarrow 4f^6$ transition is detected across the amorphization process. A small decrease in the white-line intensity is observed until the onset of amorphization (Figure~\ref{fig:fig5}(a)) and it continues across the transition. The suppressed white-line is not recovered upon pressure release (Figure~\ref{fig:fig5}(a) shows XANES spectrum at 2.0 GPa after releasing the pressure in the cell). The XMCD signal as a function of pressure is also displayed in Figure~\ref{fig:fig5} panel (b). A strong decrease in the XMCD intensity occurs at higher pressures. Note that ferromagnetic order is not recovered when the pressure in the cell is released. The small persistent XMCD signal observed after amorphization is due to the paramagnetic response of Eu$^{2+}$ ions in the applied field.

\begin{figure}[!h]
\centering
\includegraphics[trim=0.1cm 0.0cm 0.4cm 0.0cm, clip=true, totalheight=0.48\textheight, angle=0]{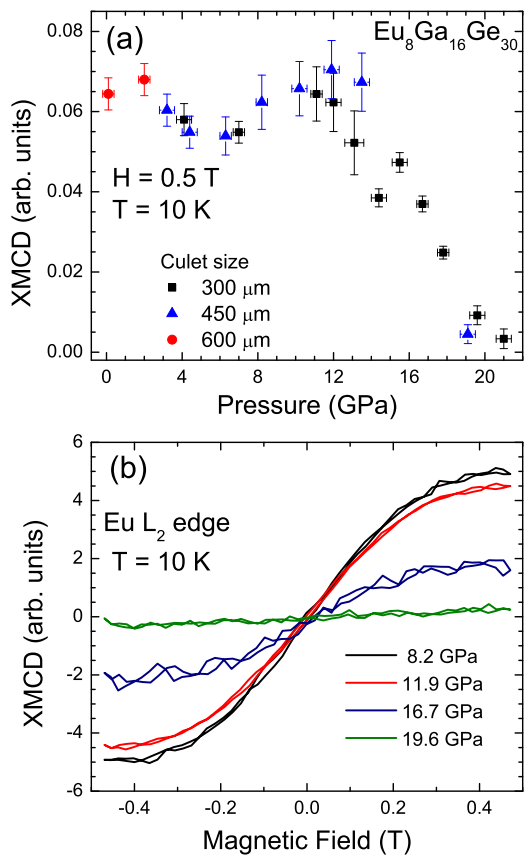}
\caption{(color online) Eu L$_2$-edge XMCD signal and hysteresis loops in type-I clathrate Eu$_8$Ga$_{16}$Ge$_{30}$ obtained at $T=10$~K for different applied pressures. The panel (a) shows the XMCD integrated intensity as a function of pressure and the panel (b) shows hysteresis loops for some selected pressure values.}
\label{fig:fig6}
\end{figure}

Figure~\ref{fig:fig6}(a) shows the integrated XMCD signal obtained at $T=10$~K and $H=0.5$~T as function of pressure. The XMCD displays a small drop in the intensity around 3~GPa, followed by an increase at 7~GPa. This small anomaly in the XMCD signal could be related to a change in the carrier concentration with pressurization, as previously hypothesized.\cite{Umeo_JPCS_2012} However, this intensity variation is roughly within the error bars. At higher pressures the XMCD intensity clearly shows a sharp suppression in the 16-19~GPa range. The fast reduction in the XMCD signal matches quite well the concomitant structural changes seen in the XRD measurements, demonstrating that the crystalline-to-amorphous transition has an adverse effect on the magnetic ordering in this compound. In addition, the hysteresis loops for selected pressures shown in Figure~\ref{fig:fig6}(b) reveal that this pressure-driven transition suppresses the ferromagnetic order. A transition to an antiferromagnetic phase is unlikely due to the amorphization of the material and lack of long range magnetic order. The abrupt XMCD collapse coupled with large hysteresis/irreversibility of structural and electronic properties indicate that the transition is first order and that the amorphization hampers the long range magnetic order of Eu$_8$Ga$_{16}$Ge$_{30}$ as a result of frustrated exchange interactions triggered by the structural disorder, resulting in a paramagnetic amorphous state.

\section{DISCUSSION}

The guest-host interactions are believed to dominate the physical response in clathrates, hence new properties are expected to emerge by inducing changes in the inter-atomic distances. The investigation of the structural, electronic and magnetic properties of the type-I clathrate Eu$_8$Ga$_{16}$Ge$_{30}$ allow us to determine changes in its ground state properties under high pressures.

Powder XRD performed at 10~K and 30~K reveal a crystalline-to-amorphous phase transition at $\sim18$~GPa where the unit cell is compressed to $\sim81\%$ of the ambient pressure volume. The evolution of XRD patterns as a function of pressure does not show any additional peaks or splittings characteristic of a change in crystal symmetry, as can be seen in Figure~\ref{fig:fig2}. A possible change in the crystallographic site of the Eu ion inside the large cage, from 6$d$ to 24$k$ would be visible by a change in intensities ratios of some intense Bragg peaks (such as the (222), (320) and (321) in Figure~\ref{fig:fig2}(a)). Such effect was not observed, suggesting that the Eu ion remains off-center (24$k$ site) in the entire pressure range. The powder patterns only reveal a reduction of intensity and a shift of the peaks to low angle due to the lattice compression. The existence of an off-center position of the Eu ion inside the larger cage, compatible with rattling phenomena, is in line with the observation of $T^*$ up to at least 11.4 GPa by resistivity measurements.\cite{Umeo_JPCS_2012}

The lattice parameter ($a$), bulk modulus (B$_0$) and amorphization pressure (P$_a$) observed for different clathrate compounds are summarized in Table~\ref{tab:table1}. Compared to other type-I clathrates investigated under pressure, Eu$_8$Ga$_{16}$Ge$_{30}$ shows the lowest threshold pressure to reach the amorphous phase. Nevertheless, its amorphization pressure is similar to those of type-III clathrates such as Ba$_{24}$Ge$_{100}$ (P$_a=20$~GPa) and Ba$_{24}$Si$_{100}$ (P$_a=23$~GPa).\cite{Miguel_HPR_2005} Such similar threshold is likely related to the presence of open cages and weak guest-host interaction observed in these cases.

\begin{table} [!hbt]

\caption{Unit cell lattice parameter (a), bulk modulus (B$_0$) and amorphization pressure (P$_a$) for different type-I clathrate compounds$^a$.}}
\begin{tabular}{c   c   c   c c c  c c c c  }
\hline
\hline

         Clathrate 		                &&         &  a (\AA)	     &&   &  B$_0$ (GPa)	     & &  &  P$_a$ (GPa) \\

\hline

\vspace{0.06cm}
Eu$_8$Ga$_{16}$Ge$_{30}$~$^{b}$	 &&	 &10.706	       &    &     &  65 $\pm$ 3      & &   & 18 $\pm$ 1    \\
\vspace{0.06cm}
Ba$_8$Ga$_{16}$Ge$_{30}$	 &	& &10.783	            &&    &  67.2                 & &  & $>$ 40    \\
\vspace{0.06cm}
Sr$_8$Ga$_{16}$Ge$_{30}$	 &	& &10.721	            &&    &  $\sim$ 130                 & &  & ?~$^{c}$   \\
\vspace{0.06cm}
Ba$_8$Si$_{46}$	                 &&	 &10.328	       &     &    &  93     &              &   & 40 $\pm$ 3    \\
\vspace{0.06cm}
Rb$_{6.15}$Si$_{46}$	   &      &	 &10.286	           & &    &  293        &         &    & 33  $\pm$ 1   \\
\hline
\hline
\label{tab:table1}

\end{tabular}
 \flushleft{ $^{a}$ Data extracted from Ref.~\onlinecite{Miguel_HPR_2005}. ~$^{b}$ This work. $^{c}$ XRD measurements were performed only up to 7~GPa and did not show any amorphization.\cite{Meng_JAP_2001}
\end{table}

Typical type-I clathrate compounds such as Ba$_8$Ga$_{16}$Ge$_{30}$ and those with the framework composed of Si atoms such as Ba$_8$Si$_{46}$ and Rb$_{6.15}$Si$_{46}$ show a transition to an amorphous phase at pressures higher than 33~GPa, despite all compounds having similar lattice parameters at ambient pressure. The clathrate Sr$_8$Ga$_{16}$Ge$_{30}$ did not show any amorphization or volume-collapse up to 7~GPa,\cite{Meng_JAP_2001} but these are likely to appear at higher pressures. The bulk modulus for the Eu$_8$Ga$_{16}$Ge$_{30}$ clathrate obtained at low temperature with the Murnaghan EOS is $65\pm3$~GPa. Comparing with other clathrates, Eu$_8$Ga$_{16}$Ge$_{30}$ and Ba$_8$Ga$_{16}$Ge$_{30}$ are the most compressible.

The guest atoms for the materials cited in Table~\ref{tab:table1} have significantly different ionic radii with Eu ions having the smallest radius (r$_{Eu}$ = 1.09~\AA; r$_{Sr}$= 1.13~\AA; r$_{Ba}$ = 1.35~\AA ~and r$_{Rb}$ = 1.48~\AA). Consequently, the larger voids in the cages and the larger rattling amplitudes in Eu$_{8}$Ga$_{16}$Ge$_{30}$ may be responsible for the lower observed pressure threshold to amorphization. In the case of Rb$_{6.15}$Si$_{46}$ the lower transition pressure as compared to the Ba$_{8}$Si$_{46}$ compound may be related to the significant fraction of empty cages due to non-stoichiometric Rb content. X-ray photoelectron spectroscopy (XPS) on type-I Ga-Ge clathrates and Sr$_{8}$Si$_{46}$ reported that the specific guest ions inside the tetrakaidecahedrons influence the cage framework structure.~\cite{Tang_PRB_2008} For the clathrate Eu$_{8}$Ga$_{16}$Ge$_{30}$ a  probable hybridization between the Eu 5$d$ states and the Ga/Ge 4$sp$ orbitals coupled with the fact that the Eu guest ion is off-center creates a shape anisotropy in the framework, which could produce a cage structure that is less mechanically stable. Consequently, for clathrates
with the on-center guest ions such as Ba$_{8}$Ga$_{16}$Ge$_{30}$, the framework is more stable leading to higher amorphization pressure. Moreover, band structure calculations performed for the A$_{8}$Ga$_{16}$Ge$_{30}$ series (with A = Ba, Sr and Eu) show a strong hybridization between the unoccupied guest ion $d$ states and the antibonding framework states.\cite{Madsen_PRB_2003, Liu_NMMA_book_2009, Blake_JCP_2001,Blake_JCP_1999} Both Eu$_{8}$Ga$_{16}$Ge$_{30}$ and Sr$_{8}$Ga$_{16}$Ge$_{30}$ clathrates were reported to have similar band gap energy although smaller when compared to the Ba$_{8}$Ga$_{16}$Ge$_{30}$ compound.\cite{Madsen_PRB_2003} The difference in the band gap is directly related to the ionic radii in which the larger element donates more effectively the electrons to the cages.\cite{Blake_JCP_2001} The higher band gap energy for the Ba clathrate can also explain the higher amorphization pressure due to changes in the stability between the $sp^3$ bonding orbitals and the antibonding. In addition, the smaller sizes for Sr and Eu guest ions and the capacity to develop the rattling motion affects the guest-framework interaction, resulting in an anisotropic network which does not happen when the ion is in the center of the cage. This confirms that the hybridization between atomic orbitals of guest ions and cage atoms is largely dependent on atom type and consequently the pressure required for the crystalline-to-amorphous transition is dependent on this guest-host interaction, the size of the guest atom and the fractional atomic occupation of the host cages. To better understand the influence of rattling motion to the observed pressure-induced amorphization, we believe that the study of Sr$_{8}$Ga$_{16}$Ge$_{30}$ is important. The guest rattling motion in this Sr clathrate is smaller than in Eu$_{8}$Ga$_{16}$Ge$_{30}$ and larger than in Ba$_{8}$Ga$_{16}$Ge$_{30}$, thus a comparative evolution of their properties under pressure should shed light on the matter.

The absorption measurements at the Eu L$_2$-edge provide further information about the Eu-cage interaction in Eu$_{8}$Ga$_{16}$Ge$_{30}$. The analysis of the XANES spectra can provide information about the density of unoccupied states above the Fermi Level and some possible valence change in the Eu ions. No indication of an Eu$^{2+}$  to Eu$^{3+}$ transition was observed in Figure~\ref{fig:fig5}(a), showing that Eu preserves its 4$f^7$ configuration across the amorphization. The L$_3$ edge XANES white-line is directly related to the number of empty 5$d$ states. Prior to amorphization, a continuous reduction of the white-line is observed indicating a small enhance in 5$d$ occupation, which seems to further increase across the amorphous transition. Note that a pressure induced 6$sp$ to 5$d$ is widely known to exist in lanthanides.\cite{Bi_PRB_2012, Duthie_PRL_1977} Thus, it is unclear if the suppression of the white line is related to an internal 6$sp \rightarrow $ 5$d$, or due a Ge/Ga 4$sp \rightarrow$ Eu 5$d$ charge transfers. The increase in 5$d$ occupation seems to contrast with previous reports of charge transfer from the guest ion to the Ga/Ge frameworks in Ba$_{8}$Ga$_{16}$Ge$_{30}$ and Sr$_{8}$Ga$_{16}$Ge$_{30}$ clathrates.\cite{Blake_JCP_2001} The 5$d$ orbital is more binding than $sp$, thus we speculate that the observed increase in 5$d$ occupation contributes to the amorphous transition by collapsing the cages. However, as discussed for the Ba$_8$Si$_{46}$ clathrate,\cite{Miguel_EL_2005} the changes in white-line intensity can be rooted in more than one mechanism, namely, changes in 5$d$ electron occupation without a change in local structure, such as a result of Eu-Ga and Eu-Ge hybridization, or a change in local structure such as relative changes in Eu-Ga and Eu-Ge distances. Either mechanism can play a role in altering the mechanical stability of the cages and lead to the crystalline-to-amorphous transition.

As mentioned earlier, the decrease in XMCD intensity as a function of pressure is not due to a valence change in Eu atoms (Eu$^{3+}$ is a $J$=0 ion), but rather to the collapse of crystalline order. This is supported by the XRD patterns where the amorphization occurs within a similar pressure range. In addition, the small XMCD signal remaining around 20~GPa (20 times smaller than at ambient pressure) is explained by the paramagnetic response of Eu$^{2+}$ ions to the 0.5~T applied magnetic field.

Finally, the lower threshold to reach the amorphous phase for the Eu$_{8}$Ga$_{16}$Ge$_{30}$ clathrate may result in important thermoelectric developments, especially since after reaching the amorphous phase the clathrate Eu$_{8}$Ga$_{16}$Ge$_{30}$ did not return to its crystalline phase even when the pressure was released and the sample heated to 300~K. This novel material in amorphous state even at room temperature may prove interesting in terms of thermoelectric performance if the electronic transport can remain manageable in order to maximize the power factor $S^2/\rho$, or it may at least provide an enlightening comparison of the thermal and electrical conductivities of the amorphous phase with the crystalline ones in this PGEC material. Thermoelectric measurements are being planned with the samples in the amorphous phase to clarify these properties.


\section{CONCLUSIONS}

We have investigated the clathrate Eu$_8$Ga$_{16}$Ge$_{30}$ under high pressures at low temperature by powder XRD and XANES/XMCD measurements. A crystalline-to-amorphous phase transition was observed above 18~GPa by XRD. This amorphization has a dramatic influence on the magnetic properties as can be observed from XMCD measurements on the Eu L$_2$ edge. The absorption measurements showed a sharp decrease in the XMCD intensity near the crystalline-amorphous transition, as a result of frustration of the exchange interactions between Eu$^{2+}$ local moments in the structurally-disordered phase. We did not observe any valency change of the Eu ions inside the cages. Both structural and magnetic measurements indicate that the structure undergoes an irreversible amorphization process with pressure; the crystalline and magnetic long range ordering are not recovered when the pressure is released in the amorphous state. As observed in other isostructural clathrate compounds, the main origin of this crystalline-to-amorphous transition is the mechanical instability of the framework~\cite{Tse_PRL_2002} and its modification under pressure due to guest-framework interactions. The novel amorphous phase might play an important role in the development of clathrates as thermoelectric materials.

\section*{acknowledgments}
This work was supported by FAPESP (SP-Brazil) under Contracts No. 2009/10264-0, 2011/24166-0, 2012/10675-2 and 2012/17562-9. Work at Argonne is supported by the U.S. Department of Energy, Office of Science, Office of Basic Energy Sciences, under Contract No. DE-AC-02-06CH11357. We would like to thank Curtis Kenny-Benson and Dmitry Popov for their assistance at 16-BMD beamline, Sergey N. Tkachev for the gas loading, and P. F. S. Rosa and W. Iwamoto for assistance in sample preparation. We also thank the GSECARS group for use of the gas loading and laser drilling facilities.





\end{document}